\documentstyle[12pt]{article}

\newcommand{\be}{\begin{equation}}
\newcommand{\ee}{\end{equation}}
\newcommand{\bal}{\begin{array}{l}}
\newcommand{\eal}{\end{array}}
\newcommand{\bea}{\begin{eqnarray}}
\newcommand{\eea}{\end{eqnarray}}

\begin{document}
\baselineskip 20pt
\begin{titlepage}
\begin{flushright}
hep-th/9610045\\
CALT-68-2079 \\
September 1996
\end{flushright}

\vskip 0.2truecm

\begin{center}
{\Large {\bf Microcanonical D-branes and Back Reaction}}
 \vskip .05cm
{\Large {\bf }}
\end{center}

\vskip 2.2cm

\begin{center}
{\bf Esko Keski-Vakkuri}$^1$ and {\bf Per Kraus}$^2$
\vskip .1cm
{\it California Institute of Technology \\
 Pasadena CA 91125, USA \\
 e-mail: esko,  perkraus@theory.caltech.edu}
\end{center}

\vskip 2.5cm

\begin{center}
{\small {\bf Abstract: }}
\end{center}
{ We compare the emission rates from excited D-branes in the 
microcanonical ensemble 
with back reaction corrected emission rates from black holes in field theory.  
In both cases,
the rates in the high energy tail of the spectrum differ markedly from 
what a canonical ensemble or free field theory approach would yield.  
Instead of being proportional to a Bose-Einstein distribution function, the
rates in the high energy tail are proportional to $e^{-\Delta S_{\rm BH}}$, 
where  $\Delta S_{\rm BH}$ is the difference in black hole
entropies before and 
after emission.  After including the new effects, we find 
agreement, at leading order, between the D-brane and field theory rates  
over the entire range of the spectrum.  
}
\rm 
\noindent
\vskip 2.0 cm

\small
\begin{flushleft}
$^1$ Work supported in part by a DOE grant DE-FG03-92-ER40701.\\
$^2$ Work supported in part by a DOE grant DE-FG03-92-ER40701 and by
a DuBridge Fellowship.
\end{flushleft}
\normalsize
\end{titlepage}

\newpage

\section{Introduction}

In the past year remarkable progress has been made in obtaining a microscopic
description of certain black holes in string theory \cite{review}.  
The most striking
results have been found in the case of black holes which can be understood,
in particular regions of moduli space, as being weakly coupled bound states
of D-branes \cite{StrVaf,CalMal,HorStr,ms,jkm,bmpv}.  Various quantities 
such as the extremal and near-extremal
entropies, and emission and absorption rates, have been computed in this
regime as an expansion in small parameters, and then extrapolated to the 
strong coupling regime where a field theory description of the black hole is
valid \cite{list,MalSus,DasMat,GubKle,MalStr}.  
Although there would appear to be no simple justification for such
an extrapolation, the results so obtained agree precisely with the predictions
of field theory.  The agreement is especially provocative considering the
very different natures of the D-brane and field theory calculations.  

In computing the emission rate a number of approximations are made, on both
the D-brane and field theory sides.  In this paper we will consider processes
for which one such approximation becomes invalid, and show how including some
previously neglected effects changes the emission rates and allows the D-brane
and field theory predictions to match in  a  non-trivial manner.  In
particular, we consider the emission of relatively high energy quanta, those
with energies comparable to the total excitation energy of the black hole 
above 
extremality.  On the D-brane side, the approximation which becomes invalid
is the use of the canonical ensemble to compute the distribution functions
for open string excitations.  In the context of the canonical ensemble the
distribution functions are given by the Bose-Einstein or Fermi-Dirac 
distributions; however, the D-brane is properly in the microcanonical
ensemble --- fixed energy --- and the distribution functions which follow
give entirely different results for the high energy tail of the spectrum. 
Given that this is the case, in order for the agreement to persist there must
be a corresponding effect on the field theory side.  We show that the effect
is the gravitational self-interaction of the emitted quanta, and that 
properly taking this 
into account leads to  a modified emission rate which agrees with the D-brane
result in the high energy tail. 

A method for including self-interaction effects was developed in \cite{KraWil}.
(See also \cite{BalVer}.)
If attention is restricted to the s-wave, then the gravitational field by
itself has no dynamical degrees of freedom, and it becomes sensible to
integrate it out. This can be done by working in the Hamiltonian formalism
and solving constraints.  What is left is an effective action for the 
remaining matter degrees of freedom, which can then be used to obtain
a  corrected field equation.  In contrast to the standard approach of 
quantizing a field on a fixed background, this approach allows the geometry
to change in response to the matter, and enforces energy conservation 
for the complete gravity plus matter system.  This distinction is directly
analogous to that between the canonical and microcanonical ensembles, in that
the energy is only held fixed in the latter case.

Although our calculations will go through more generally, for definiteness 
we consider the specific D-brane configuration studied in  \cite{CalMal} and
 \cite{HorStr}.   The black hole is realized as a bound state of $Q_1$ 
D1-branes, $Q_5$ D5-branes, and total left moving momentum number $N$, all 
wrapped around  $S^1\times T^4$.  The size of the $S_1$ is $L=2\pi R$, and
that of $T^4$ is $V$.  In addition, we adopt the model of \cite{MalSus} in 
which the open strings attached to the D-brane effectively behave as though 
they were moving on a single D1-brane wrapped $Q_1 Q_5$ times around $S_1$,
with a total length $L'=2\pi Q_1 Q_5 R$.  In this picture, the total left
moving momentum is given in terms of $N'$ as  $P = N/R = N'/(Q_1 Q_5 R)$.  
If only left moving momentum is present the black hole is extremal and BPS
saturated.  The near-extremal hole is obtained by adding left and right moving
momentum, while keeping $N'=N_L'-N_R'$ fixed.  
The entropy\footnote{We will be taking the term `black hole entropy' and the 
symbol
$S_{\rm BH}$ to mean $A/(4G_N)$.  The true entropy of the black hole is 
equal to this plus corrections.}
 of the black hole is given by
\be
S_{\rm BH} \ = \ 2\pi ( \sqrt{N'_L}+\sqrt{N'_R\,}\,).
\label{entropy}
\ee
We will only be considering the dilute gas regime \cite{MalStr}, so that the
contribution to the entropy from anti-branes is suppressed.  
The momentum is carried by massless excitations of
open strings connecting the D1-brane to the D5-brane.  These 
excitations correspond to a one dimensional gas of four species of massless
bosons and fermions confined to a box of length $L'$.  The average number
of bosonic quanta of a particular species and with energy $\omega_k$ is given by 
the  distribution functions $\rho_L(\omega_k), \ \rho_R (\omega_k)$.  

The non-extremal configuration will decay
due to the collision of left and right moving excitations.  The resulting
rate for the emission of a single species of neutral scalars was computed
in \cite{DasMat}:
\begin{equation}
 \Gamma_D (\omega_k) =  \ \frac{\kappa_5^2 L}{4} \omega_k
              \  \rho_L (\omega_k/2) \rho_R (\omega_k/2) \ \frac{d^4k}{(2\pi )^4}
\end{equation}
On the other hand, the black hole decay rate is
\begin{equation}
 \Gamma_H (\omega_k) = \sigma_{{\rm abs}} (\omega_k) \ \rho_H (\omega_k) 
                \ \frac{d^4k}{(2\pi)^4} 
\label{bhrate}
\end{equation}
where $\sigma_{{\rm abs}} (\omega_k)$ is the grey body factor, equal to the 
classical absorption cross section,  and $\rho_H (\omega_k)$ is the Hawking 
amplitude for particle creation by a black hole.  If we follow \cite{DasMat}
and restrict ourselves to the near-extremal region\footnote{we are required
to be sufficiently near extremality that the wavelengths of emitted quanta are
much greater than the Schwarzschild radius, $R_s$.  Since the energy of an
emitted quantum can be at most $4\pi N'_R/L'$, we will require that 
$N'_R \ll L'/R_s$.}  and also take 
$N_L' \gg N_R' \gg 1$,
then these rates reduce to:
\be
\Gamma_D (\omega_k) = \ A_H\, \rho_R (\omega_k/2) \ \frac{d^4k}{(2\pi )^4}
\quad\quad ; \quad\quad
\Gamma_H (\omega_k) =  \  A_H\, \rho_H (\omega_k) \frac{d^4k}{(2\pi )^4}
\label{simplerate}
\ee
The two rates thus agree provided $\rho_R (\omega_k/2)=\rho_H (\omega_k)$.  For 
sufficiently small $\omega_k$ the canonical ensemble can be used to compute
$\rho_R (\omega_k/2)$, and self-interaction effects can be ignored in computing
$\rho_H (\omega_k)$.  The distribution functions then each 
take the Bose-Einstein
form, with the temperature given by the Hawking temperature, and the rates 
agree.  However, these approximations fail for larger $\omega_k$ and additional
analysis is required to demonstrate agreement.

The remainder of this paper is organized as follows.  In section 2 we 
study distribution functions in the microcanonical ensemble.  For the extreme
high energy tail, {\em i.e.} for quanta with energies equal to the total 
energy of the system, simple reasoning leads to the result that the average
number of quanta present is $e^{-S}$, where $S$ is the microcanonical
entropy of the system.
This is so because there is only one state, out of $e^{S}$ total states, for
which all of the energy is concentrated  in a single quantum.  For lower
energies we perform a saddle point calculation to derive the leading
corrections to the canonical distribution function.  In section 3 we turn to
the field theory calculation of $\rho_H (\omega_k)$ in the presence of 
self-interaction effects,  using a refined version of the 
techniques developed in \cite{KraWil}.  The calculation is particularly
clean in the case where the black hole decays to the extremal state by emitting
only a single quantum, for then multi-particle interaction effects can
be ignored.  Remarkably, the amplitude turns out to be 
simply $e^{-\Delta S_{\rm BH}}$,
where $\Delta S_{\rm BH}$ is the difference in 
entropies of the black hole  before and
after emission.  This form holds for a general spherically symmetric black
hole.  We therefore find agreement at leading order between the D-brane and 
field theory 
emission rates in the high energy tail.  In section four we further discuss
our results, and consider how agreement beyond lowest order might be reached
by a more detailed field theory analysis.

\section{Boltzmann Factors from Microcanonical Ensemble}

 In the `fat' black hole model of  \cite{MalSus}, 
the rightmoving excitations on the multiply wound D1-brane 
consist of 4 species of massless bosonic and fermionic
open string excitations. In what follows, we shall consider an arbitrary
number $f$ of species; in the end one can set $f=4$.
Half of the excess energy above extremality is  distributed
among the rightmoving quanta, which have  momenta  quantized in 
units of $p_R = 2\pi / L'$.  The total momentum of the rightmovers must add 
up to $P_R = 2\pi N'_R / L'$.  Thus the counting problem consists roughly 
of adding up positive integers to get a total of $N'_R$.  

For a system of $f$ superconformal fields,  
the number of states consistent with a total level number $n$ is given by
(see {\em e.g.} \cite{GSW}): 
\be
     d_{n} \equiv e^S \approx n^{-(f+3)/4} \ \exp \{ \pi \sqrt{f n} \} \ .
\label{d_n}
\ee
We would now like to compute the microcanonical Boltzmann factor, $P(n_r=k)$, 
which is the probability  to find  $k$ 
bosonic quanta of a given species in a state with energy $2\pi r/ L'$.   
 $P(n_r=k)$ is proportional to the number, $d(n_r=k)$, of such states. 
If there are $k$ quanta at
level $r$, they contribute $kr$ to the total level number, leaving $n-kr$ 
units of momentum to be distributed among the remaining levels.  
 $d(n_r=k)$ is thus equal to the number of states with total level number
$n-kr$ for a system of $f$ species of bosons and 
fermions, with one bosonic level $r$ removed.  Then, the Boltzmann factor is
given by
\be
      P(n_r=k) = \frac{1}{{\cal N}} \ d(n_r=k)
\ee
where $1/{\cal N}$ is a normalization factor such that $\sum_k P(n_r=k)=1$. 
The distribution function is defined to be the average number of quanta of a 
given bosonic species in level r:
\be
\rho(r) = \sum_k P(n_r=k)\, k.
\ee

The simplest case to consider is $r=n$.  There is clearly only one such state,
so $d(n_n=1)=1$.  Since $d(n_n=0)=d_n-1$, we find
\be
\rho(r=n) = 1/d_n.
\label{easyrho}
\ee

For $r < n$ it is not possible to proceed so simply, but over most of the 
spectrum we can compute perturbatively and find the leading corrections to 
the values given by the canonical ensemble.  
The counting is done in the usual manner, by using the 
partition function as a generating function for the level degeneracies
and then projecting out the individual factors \cite{GSW}. In our case, the
appropriate partition function for the system with one level $r$ removed is:
\be
    G_r(w) = (1-w^r)\ \prod^{\infty}_{s=1} 
                   \left( \frac{1+w^s}{1-w^s} \right)^f  \ .
\ee
Asymptotically, as $w\rightarrow 1$,
\be
    G_r (w) \approx (1-w^r) \ \left( \frac{\ln w}{-2\pi} \right)^{f/2} 
                 \ \exp \left(\frac{-f\pi^2}{4 \ln w} \right) \ .   
\label{asymp}
\ee
The degeneracy is given by
\be
     d(n_r=k) = \frac{1}{2\pi i}  \oint  \frac{dw}{w^{ n+1-rk}} \ G_r(w) \ .
\label{contour}
\ee
The integral will be approximated by the saddle point method,
using the asymptotic formula (\ref{asymp}) for $G_r$. For this to be accurate 
we require that
$$
    \ n - rk \gg 1 \ .
$$

If we are interested in finding the microcanonical Boltzmann
factor up to leading corrections away from its canonical form, it is 
appropriate to identify the relevant small parameters.  It will be convenient
to perform expansions in 
$$ 
   \frac{1}{\sqrt{n}} \ \ {\rm and} \ \ \frac{r}{n} \ ,
$$
We skip the tedious details and simply state the result:
\be
 d(n_r=k) \sim \exp \left\{ -\frac{\pi \sqrt{f}kr}{2\ n^{1/2} }
                  -\frac{\pi \sqrt{f}(kr)^2}{8\ n^{3/2}}
                  +\frac{(f+3)kr}{4\ n}
                  +\frac{\pi}{4}\frac{\sqrt{f}}{e^{\pi r\sqrt{f / 4n }}-1}
                   \frac{kr^2}{n^{3/2}} + \ldots \right\} \ .
\label{degen}
\ee
We have omitted writing the $k$ independent factors, as these will 
just drop out when we compute $P(n_r=k)$.  
The first term in the exponent is the 
usual thermal canonical ensemble result, and 
the next three terms give the leading microcanonical corrections.
These corrections have different origins in the saddle point calculation.
The first correction term has the same origin as the canonical term. The
second correction term comes from two different contributions: The term
with the coefficient $f/4$ can be traced back to the term
$$
        \left( \frac{\ln w}{-2\pi} \right)^{f/2}
$$
in the partition function (\ref{asymp}), and the term with 
coefficient $3/4$ comes from the square root 
prefactor in the saddle point evaluation.
The last correction term represents the effect of the removed bosonic
level $r$ in the partition function. The omitted terms $+\ldots$ in
the exponent are subleading corrections.

It is useful to distinguish the following parameter regions:
\begin{enumerate}
\item  In the low energy part of the spectrum, $r  \ll n$, the degeneracy 
reduces to the canonical result, but with a corrected temperature coming from
the third and fourth terms.  
\item If $r \sim \sqrt{n}$, the two expansion parameters are of the same 
order of magnitude, and the different microcanonical corrections 
contribute roughly equally.  This region corresponds to
the energy of the level being of the order of the temperature. 
\item For the higher levels, $n \gg r \gg \sqrt{n}$, the second 
term gives the leading correction.  
\item When $r \sim n$, the saddle point method
is no longer  valid.
\end{enumerate}

Let us focus on region 3.  In this region, up to the accuracy considered, we
can write
\be
 d(n_r=k) \ \approx \ \exp{\left[\pi\sqrt{fn}-\pi\sqrt{f(n-kr)}\,\right]}.
\ee
Furthermore, in computing  $P(n_r=k)$ it is accurate to
replace  $\sum_k$ by the $k=0$ term, and for  $\rho(r)$  by the  $k=1$ term.  
Therefore, we find
\be
\rho(r) \ \approx \ \exp{\left[\pi\sqrt{f(n-r)}-\pi\sqrt{fn}\,\right]}.
\label{hightail}
\ee

To translate this back into language appropriate for the black hole, we 
 use $\omega_k/2 = 2\pi r/L'$ and  the expression for the black hole
entropy (\ref{entropy}), and set $f=4$, $n=N'_R$.  In region 3, our result 
then takes the form\footnote{The momentum of the left movers, $N'_L$, can be 
thought of as 
a constant, as the change only gives a subleading correction.}
\be
\rho_R(\omega_k/2) \ \approx \ 
\exp{\left[S_{\rm BH}(M-\omega_k)-S_{\rm BH}(M)\right]},
\ee
where $M$ is the mass of the black hole.  
 This formula also gives the leading result in the extreme
high energy tail, when $r=n$.  However, there we also know the more accurate
result,
\be
\rho_R(\omega_k/2) \ \approx \ 
\left(\frac{S_{\rm BH}(M)-S_{\rm BH}(M-\omega_k)}{2\pi}\right)^{7/2}
\exp{\left[S_{\rm BH}(M-\omega_k)-S_{\rm BH}(M)\right]},
\label{microrate}
\ee
where we used (\ref{easyrho}), (\ref{d_n}) and (\ref{entropy}). 
Note that (\ref{microrate})
contains all terms of (\ref{degen}) except those arising from removing
the $r$th bosonic level in the partition function.  

We have found that the correct microcanonical description of the
gas of open string excitations on the D-string leads to a prediction:
there are corrections to the exactly thermal behavior, and presumably
these should correspond
to some new effects on the field theory side. Since the corrections
were seen to have different mathematical
origins, it is possible that the corresponding field
theory corrections are associated with a variety of different physical 
effects. 
Furthermore, the result is partially dependent on the specific D-brane
model.
For example, in (\ref{d_n}) the prefactor contains
an explicit dependence on the number of species $f$ of superconformal 
fields --- $f$ does not appear in the combination $fn$ as in the exponential
term --- and this dependence shows up as the $7/2$ power in  
(\ref{microrate}). These subleading
terms would be different  in the
model  appropriate to the black string limit, where
there are $Q_1 \ (Q_5)$ singly wound D1 (D5) -branes, and the number of  
species is $f=4Q_1Q_5$ instead of $4$. 
However,  the same leading exponential 
terms in (\ref{d_n}), (\ref{microrate}) would still appear. 
In the next section we shall show that the leading term in the high energy
tail region can be computed in field theory, provided gravitational
self-interaction effects are properly included.

\section{Emission Rate From Field Theory}
\subsection{Metric}

In this section we will be using a slightly non-standard set of coordinates
for the black hole metric \cite{kw}:
\be
ds^2~=~-[N_{t}(r)~dt]^2~+~[dr+N_{r}(r)~dt]^2~+~r^2~d\Omega_{d-2}^2.
\ee
The metric can always be put in this form, provided that the geometry is 
spherically symmetric and has a Killing vector which is timelike outside the
horizon.  The advantage of these coordinates is that they are well behaved
at the horizon.  For example, for a four dimensional Reissner-Nordstr\"{o}m
solution: $N_t=1~,~N_r=\sqrt{2M/r-Q^2/r^2}$.  
  
Since the equation for null
geodesics is $\dot{r}=\pm N_t-N_r$, with the plus(minus) sign applying for
outgoing(ingoing) trajectories, we see that the horizon, $r=R$, is 
determined from the condition $N_{t}(R)-N_{r}(R)=0$.  
Near the horizon $N_t-N_r$ behaves as
\be
N_{t}(r)-N_{r}(r) ~\sim~  (r-R)\kappa~+~ {\rm O}\!\left((r-R)^2\right) 
\ee
where $\kappa$ can easily be seen to be the surface gravity of the black hole. 
For comparison with other forms of the metric, it is useful to note that a 
change of time coordinate brings the metric into the form
\be
ds^2~=~-[N_t^2(r)-N_r^2(r)]d\tilde{t}^2~+~\frac{N_t^2(r)}{N_t^2(r)-N_r^2(r)}
dr^2~+~r^2~d\Omega_{d-2}^2.
\ee
$N_t$, $N_r$ are generally functions of various charges as well as the mass of
the black hole;  in what follows we only show the mass dependence explicitly.
We thus write: $N_t(r;M)$, $N_r(r;M),R(M)$,$\kappa(M)$.

\subsection{Black hole radiance including self-interaction}

We now turn to the computation of emission probabilities in field theory, 
and include effects due to self-interaction.  Only the s-wave modes are
considered, as these dominate the emission process at low energies.  
Black hole radiance results from the mismatch between the two natural vacuum
states which arise in the quantization of a field propagating on a black 
hole spacetime.  The first vacuum state to consider is  the one most naturally
employed by an asymptotic observer.  Such an observer would expand the field
operator in terms of mode solutions which have the time dependence 
$\exp{(-i\omega t)}$, where t is the Killing time:
\be
\phi(t,r)~=~\int \!dk\, [a_k u_k(r)e^{-i\omega t}~+~a_k^{\dagger}u_k^*(r)
e^{i\omega t}].
\ee
For large $r$, $u_k(r)\rightarrow e^{ikr}/r^{(d-2)/2}$.
$a_k$ and $a_k^{\dagger}$ destroy and create quanta of definite energy,
and the corresponding vacuum state $\left|0_u\right>$ is defined by
$a_k\left|0_u\right>=0$.  However,  the modes $u_k(r)$ become singular at the 
horizon,
and the state $\left|0_u\right>$ yields an infinite result for the 
energy-momentum density measured by a freely falling observer crossing the
horizon.  Thus if physics is to be well behaved at the horizon, the state of 
the field resulting from black hole formation by collapsing matter cannot be
$\left|0_u\right>$.
A state which {\em is} well behaved at the horizon can be obtained by expanding
the field in terms of modes $v_k(t,r)$ which are nonsingular there,
\be
\phi(t,r)~=~\int \!dk\, [b_k v_k(t,r)~+~b_k^{\dagger}v_k^*(t,r)].
\ee
 Then, the state  $\left|0_v\right>$ determined by $b_k\left|0_v\right>=0$
results in a finite energy-momentum density at the horizon, and so is a 
viable candidate. 

The difference between the states $\left|0_u\right>$ and $\left|0_v\right>$
is characterized by the fact that the modes $v_k(t,r)$ contain both positive
and negative frequency components.  Defining\footnote{The precise location
of $r_f$ in the integrals  does not matter, provided it is outside the horizon
where $u_k(r)$ breaks down.  Also, note that our definitions are not quite
identical to the standard Bogoliubov coefficients, due to the absence of 
$u_k(r_f)$ factors; these factors will just cancel out in the ratio 
$|\beta/\alpha|$ and so have been omitted from the start.}
\be
\alpha_{kk'}~=~\int_{-\infty}^{\infty}\!dt\,e^{i\omega_k t}\,v_{k'}(t,r_f)
\quad\quad ; \quad\quad \beta_{kk'}~=~\int_{-\infty}^{\infty}\!dt\,e^{-i\omega_k t}
\,v_{k'}(t,r_f),
\ee
the standard result is that in the state $\left|0_v\right>$, the probability 
per unit time to emit particles with energy in the range $\omega_k$
to $\omega_k+d\omega_k$ is controlled by 
\footnote {This is true provided 
$\beta_{kk'}/\alpha_{kk'}$ is independent of $k'$, as will seen to be the case 
throughout this paper.} $|\beta_{kk'}/\alpha_{kk'}|^2$.
If the emissions are 
uncorrelated, as is the case in free field theory, then the total flux of
outgoing particles takes the form of (\ref{bhrate}) with
\be
\rho_H(\omega_k) \ = \ 
\frac{|\beta_{kk'}/\alpha_{kk'}|^2}{1-|\beta_{kk'}/\alpha_{kk'}|^2}.
\label{freerate}
\ee
On the other hand, if we consider sufficiently large $\omega_k$ such that at most
one particle can be emitted, then
\be
\rho_H(\omega_k) \ = \ 
\frac{|\beta_{kk'}/\alpha_{kk'}|^2}{1+|\beta_{kk'}/\alpha_{kk'}|^2}
\ \approx \ |\beta_{kk'}/\alpha_{kk'}|^2.
\label{singlerate}
\ee
In the latter case the emission probability is low, 
$|\beta_{kk'}/\alpha_{kk'}|^2 \ll 1$.  
As an  example, for a free field propagating on the 
four dimensional Schwarzschild
metric, it is conventional to take $v_{k'}(t,r)=e^{ik'U}$, where $U$ is the
Kruskal coordinate.  Such a mode has $t$ dependence $\exp{(-ik'e^{-t/4M})}$
leading 
to $|\beta_{kk'}/\alpha_{kk'}|^2=e^{-8\pi M\omega_k}$. From (\ref{freerate}), 
the outgoing
flux is then that of a thermal body (with grey-body factor) at the Hawking
temperature $T_H=1/(8\pi  M)$.

In the free field approximation the field obeys $\Box \phi=0$; now we would 
like to consider how the field equation is altered due to gravitational
self-interaction.  The strategy will be to obtain a corrected field equation
and then to use the corresponding mode solutions to calculate $|\beta/\alpha|$.
To proceed, we will quantize a massless, gravitating, spherical shell
surrounding the black hole.  Such an object has a position coordinate $r$,
canonical momentum $p$, and Hamiltonian $H(r,p)$.  By deriving the explicit
form for $H(r,p)$, and turning classical quantities into quantum mechanical
operators, we can derive the modified field equation.  In \cite{KraWil} this 
procedure was carried out systematically starting from the full action for the
gravity plus shell system.  The reduced Hamiltonian for the shell variables
alone was obtained by solving the gravitational constraints and inserting 
the solutions back into the action.  Here we will use a shortcut to arrive
at the same result.  The key point, which emerges from the analysis in
\cite{KraWil}, is that for a black hole of mass $M$ and shell energy $H$, the
classical trajectory of the shell is a null geodesic in the metric
\be
ds^2~=~-[N_{t}(r;M+H)~dt]^2~+~[dr+N_{r}(r;M+H)~dt]^2,
\ee
that is,
\be
\dot{r}(r)~=~N_t(r;M+H)-N_r(r;M+H).
\ee
On the other hand, this trajectory must also follow from Hamilton's equations
applied to $H(r,p)$: $\dot{r}=\partial H/\partial p$.  So 
$\partial H/\partial p = N_t(r;M+H)-N_r(r;M+H)$, and we thus find
\be
p(r,H)~=~\int_{0}^{H}\frac{dH'}{N_t(r;M+H')-N_r(r;M+H')}.
\label{pdef}
\ee
The choice for the lower limit of integration can be justified by comparing
with the free field limit, or by comparison with \cite{KraWil}.  In general,
it is not possible to invert $p(r,H)$ to obtain $H(r,p)$; fortunately, such
an expression will not be needed in what follows.  

To pass to the quantum theory we would like to make the substitutions 
\be
p \rightarrow -i\frac{\partial}{\partial r} \quad\quad ; \quad\quad
H \rightarrow i\frac{\partial}{\partial t} 
\ee
and arrive at a differential equation for the field.  However, one is met with
factor ordering ambiguities, and at best one obtains a rather unwieldy 
non-local equation.   Fortunately, the mode solutions we are interested in,
the $v_k(t,r)$, are accurately described by the WKB approximation and are
insensitive to these issues.  This is because of the large redshift involved
--- the bulk of the emission from the black hole is governed by modes which 
have a very short wavelength at the horizon, and provided the curvature at the 
horizon does not blow up, the WKB approximation is valid for such modes.  
We therefore write
\be
v_{k'}(t,r)~=~e^{iS_{k'}(t,r)}
\ee
where $S_{k'}(t,r)$ satisfies the Hamilton-Jacobi equation,
\be
\label{HJ}
\frac{\partial S_{k'}}{\partial r}~=~p(r,-\partial S_{k'}/ \partial t).
\ee
As a boundary condition, it is convenient to take
\be
S_{k'}(0,r)~=~k'r.
\ee
The solution to the Hamilton-Jacobi equation, (\ref{HJ}), is given by the
classical action. That is, if $r(t)$  is a classical trajectory then
\be
S_{k'}\left(t,r(t)\right)~=~k'r_0~+~\int_{0}^{t}\!dt\,\left[p\left(r(t)\right)
\dot{r}-H(r_0)\right],
\label{action}
\ee
where $r_0 = r(0)$, and the conserved energy $H(r_0)$ of the trajectory should
be chosen such that the boundary condition  $p(r_0)=k'$ is satisfied.  One can
check that $\partial S / \partial r = p$ and $\partial S / \partial t = -H$
as desired.

In \cite{KraWil} this construction was used to derive an explicit form for 
$v_{k'}(t,r)$, but here we will proceed in a slightly more indirect fashion.
We want to compute
\be
\alpha_{kk'}~=~\int_{-\infty}^{\infty}\!dt\,e^{i\omega_k t}\,e^{iS_{k'}(t,r_f)}
\quad\quad ; \quad\quad \beta_{kk'}~=~\int_{-\infty}^{\infty}\!dt\,e^{-i\omega_k t}
\,e^{iS_{k'}(t,r_f)}.
\ee
For large $k$ this can be done by saddle point evaluation.  The saddle point
is found from 
\be
\pm  \omega_k + \frac{\partial S_{k'}}{\partial t}~=~0,
\ee
the upper(lower) sign applying for $\alpha_{kk'}$($\beta_{kk'}$).  For
$\alpha_{kk'}$ the condition can be written as $H=\omega_k$, which simply says
 that the saddle point trajectory has energy $\omega_k$.  For $\beta_{kk'}$
the condition is instead  $H=-\omega_k$ so that the trajectory has negative
energy.  Calling these trajectories $r_{+}(t)$ and $r_{-}(t)$, we can
use the formula  for the action given in (\ref{action}) to obtain expressions
for $\alpha_{kk'}$ and $\beta_{kk'}$.  We will approximate the integrals
by the value of the integrand at the saddle point.  Of course, there is also
a prefactor coming from the second derivative at the saddle point; this 
prefactor will be ignored  for now.  
Substituting in, we find
\be
|\alpha_{kk'}|~=~e^{-{\rm Im}\int_{r_{+ 0}}^{r_f} p_{+}(r)\,dr}
\quad\quad ; \quad\quad 
|\beta_{kk'}|~=~e^{-{\rm Im}\int_{r_{- 0}}^{r_f} p_{-}(r)\,dr}
\label{tunneling}
\ee
We have not included the $k'r_0$ terms because, as we will see momentarily,
$r_0$  is always real.  Note also that the $\omega_{k} t$ term cancels with
the $\int H dt$ term in both cases.  Now, from (\ref{pdef}),
\be
p_{\pm}(r)~=~\int_{0}^{\pm\omega_k}\frac{dH'}{N_t(r;M+H')-N_r(r;M+H')}
\ee
We can find $r_{\pm 0}$ by satisfying the boundary condition
$p(r_0)=k'$.  For large $k'$,  $r_{\pm 0}$ must be very close to the 
horizon so that the redshift can convert the large valuse of $k' $ into a
 finite value for the energy of the trajectory.  Therefore,  
\be
r_{\pm 0}~=~R(M\pm\omega_k)\pm\epsilon_{\pm}
\ee
where $\epsilon_{\pm}$  are small and positive.
$r_{- 0}$ is found to be inside the horizon because of the fact that the
trajectory  $r_{-}(t)$ has negative energy.  Now, interchanging orders
of integration gives
\be
{\rm Im}\int_{r_{+ 0}}^{r_f} p_{+}(r)\,dr~=~
{\rm Im}\int_{0}^{\omega_k}\!dH'\,
\int_{r_{+ 0}}^{r_f}\frac{dr}{N_t(r;M+H')-N_r(r;M+H')}.
\ee
The imaginary part vanishes, since in the region considered $r$  is always 
outside the horizon: $N_t-N_r  > 0$. Thus $|\alpha_{kk'}|=1$.

The situation is different for $\beta_{kk'}$,
\be
{\rm Im}\int_{r_{- 0}}^{r_f} p_{-}(r)\,dr~=~
{\rm Im}\int_{0}^{-\omega_k}\!dH'\,
\int_{r_{- 0}}^{r_f}\frac{dr}{N_t(r;M+H')-N_r(r;M+H')},
\ee
since now the $r$ integration region crosses the horizon for all values of
$H'$.  The imaginary part comes from integrating over the pole where,
\be
\frac{1}{N_t(r;M+H')-N_r(r;M+H')} ~\sim ~
\frac{1}{\kappa(M+H')}\,\frac{1}{r-R(M+H')}.
\ee
Therefore \footnote {The sign of the imaginary part is chosen by comparing
with the free field limit.}
\be
{\rm Im}\int_{r_{- 0}}^{r_f} p_{+}(r)\,dr~=~
-\pi \int_{0}^{-\omega_k} \frac{dH'}{\kappa(M+H')}.
\ee
Recalling the first law of thermodynamics, $dM=(\kappa/2\pi)dS_{\rm BH}$,
we now see that
\be
{\rm Im}\int_{r_{+ 0}}^{r_f} p_{+}(r)\,dr~=~
\frac{1}{2}[S_{\rm BH}(M)-S_{\rm BH}(M-\omega_k)].
\ee
So
\be
|\beta_{kk'}|^2~=~\exp{\left[S_{\rm BH}(M-\omega_k)-S_{\rm BH}(M)\right]}.
\label{result}
\ee
For small $\omega_k$ the term in the exponent can accurately be expanded
to first order, and it is also appropriate to use the form (\ref{freerate})
since interaction effects are small.  Thus
\be
\rho_H(\omega_k) \ \approx \ \frac{1}{e^{\omega_k/T_H}-1},
\ee
where $T_H = \kappa/(2\pi)$ is the Hawking temperature.  
But for large $\omega_k$ the exponent should not be expanded, and the form
(\ref{singlerate}) should be used,
\be
\rho_H(\omega_k) \ \approx \ 
\exp{\left[S_{\rm BH}(M-\omega_k)-S_{\rm BH}(M)\right]}.
\label{fieldresult}
\ee

We would now like to make some comments about the preceeding derivation.  
First, the results presented here correct a calculational 
error in \cite{KraWil}.  Second, an interesting feature is the form of 
$\alpha_{kk'},  \ \beta_{kk'}$ given in (\ref{tunneling}), as these expressions
are the standard form for a WKB transmission coefficient in quantum
mechanics.  Thus, as was noted in \cite{KraWil}, the emission is 
naturally interepreted as being due to particles tunneling across the 
horizon.  Finally, it is remarkable that we never had to make reference to
a specific black hole geometry, so that (\ref{result}) holds for any for
any spherically symmetric black hole.  Although this feature proved
convenient here, it would also be interesting to extend the analysis to 
include effects which do depend on more details of the geometry.  

\section{Discussion}

We would now like to compare the D-brane and field theory predictions.  At
very low energies, both $\rho_R$ and $\rho_H$ go smoothly over to a 
Bose-Einstein distribution.  At higher energies, the D-brane result receives
corrections from the various terms in (\ref{degen}).  The third and fourth
terms in (\ref{degen}) correspond to corrections which are not evident in our
field theory analysis; we will speculate on their possible orgin 
momentarily.  As the energy is increased further, the leading order
distribution functions tend to their values in the high energy tail,
$\rho_R=\rho_H=e^{-\Delta S_{\rm BH}}$.  Again, the D-brane result supplies an
additional correction from the prefactor in (\ref{microrate}).  Thus,
while the two pictures give the same results to leading order, the D-brane
analysis supplies additional information.  

In making the comparison, it is important to distinguish those results which
are sensitive to the details of the particular D-brane configuration and
black hole geometry, from those which only rely on more general 
considerations \footnote{We thank S.~Mathur for a discussion
on these issues.}.
For instance, the result that the distribution functions tend, at leading 
order, to $e^{-\Delta S_{\rm BH}}$ is of the latter type. As long as there
exists a microscopic  description of the degrees of freedom, the same counting
argument that we invoked  for the average occupation number will go through.
An analogous situation holds on the field theory side in that we never made
explicit reference to the precise form of the metric.  While we are thus not
testing the D-brane model in the same way that, for instance, a grey-body 
factor calculation does, we {\em are} testing the ability of field theory to 
reproduce the results of a general microscopic description.  Such issues are
key to understanding the issues related to the apparent information loss
that occurs when the field theory description is taken literally.  That is,
assuming that the
microscopic description really is supplying a unitary S-matrix, one wants to
know which features field theory is incapable of reproducing. 

We have also discussed corrections coming from the D-brane side which are 
sensitive to the specifics of the model.  For example, the prefactor in
(\ref{microrate}) is raised to power which depends on the number of species
of bosons and fermions moving on the D-string.  
It is interesting to consider whether a more detailed field theory calculation
 would be 
able to account for these sorts of corrections.  With this in mind, let
us mention some of the effects which were neglected in our analysis.  First,
we considered only self-interactions and ignored interactions between
different emitted particles.  One might expect that such  effects can  
become appreciable in the lower energy part of the spectrum, where several 
particles can be emitted within a short time interval.  Second, there are
corrections to the WKB approximation, and to the saddle point method used
to compute $\beta_{kk'}/\alpha_{kk'}$.  We do not know how to compute 
 these corrections, but presumably they are under control for a large,
semiclassical looking black hole.  It would be interesting to examine
these issues further, and to study other processes such as scattering.  
Finally, it is very encouraging that 
the study of black holes from a string theory perspective has 
reached the stage where predictions can easily be made which are very
nontrivial to verify from a field theory point of view.  

\bigskip

{\Large {\bf Acknowledgements}}

\bigskip

We would like to thank David Lowe and Samir Mathur for helpful discussions. 

\bigskip


\end{document}